\documentstyle{article}
\def\e{\eta}
\def\dev{\partial}
\def\G{\Gamma}
\def\m{\mu}

\def\t{\tau}
\begin{document}
\pagestyle{empty}
\title{\bf The radial gauge  propagators in quantum gravity.\footnotemark
\footnotetext{Work partially supported by M.U.R.S.T.}
}
\large
\author{P. Menotti, G. Modanese\\
Dipartimento di Fisica della Universit\`a, Pisa 56100, Italy and\\
INFN, Sezione di Pisa\\
\\
D. Seminara\\
Scuola Normale Superiore, Pisa 56100, Italy and\\
INFN, Sezione di Pisa\\
}

\maketitle
\newpage
\pagestyle{plain}
\begin{abstract}
We give a general procedure for extracting the propagators in gauge
theories in presence of a sharp gauge fixing and we apply it to derive
the propagators in quantum gravity in the radial gauge, both in the first
and in the second order formalism in any space-time dimension. In the three
dimensional case such propagators vanish except for singular collinear
contributions, in agreement with the absence of propagating gravitons.
\end{abstract}

\section{Introduction.}

\indent The radial gauge (or Poincar\'e gauge) was introduced by Fock and
Schwinger
\cite{FS,Sch} in the context
of quantum electrodynamics and often used in problems connected with QCD at
the non perturbative level \cite{QCD}. The main feature of such
a gauge is the possibility of expressing correlation
functions in terms of v.e.v. of physical fields like the field strength
$F_{\mu\nu}$.
Recently an extension of the radial gauge to general relativity has been
given by
Modanese and Toller \cite {M.T.}. Here the meaning of the gauge is more
profound than in the simpler case of gauge
theories, because it involves also the reparametrization group. In fact the
radial
 coordinates in which such a gauge is realized have a well defined physical
meaning
as geodesic  or Riemann-Eisenhart coordinates that radiate from a given
origin.
The hamiltonian approach in Riemann normal coordinates has been developed
by Nelson and  Regge \cite{N.R.}. Here we shall look at the problem from the
lagrangian point of view, i.e. starting from the functional integral.
In the functional integral approach the correlation functions are computed
by averaging the products of fields like ${\cal O}(x) {\cal O}(y)$ on all
geometries weighted by the exponential of the gravitational action.
 Fixing the gauge to the radial gauge gives $x,y,\dots$ a well defined
meaning as
the points that acquire geodesic coordinates $x,y,\dots$  in each of the
geometries we are summing over.\par

\indent The problem of computing physical correlation functions has become
recently
acute when people started performing Monte Carlo  simulations in general
relativity models
\cite{reticoli}. In the
usually adopted method of the Regge calculus, one resorts to computing
expectation
values of products of fields defined on the sample geometry at a given
geodesic distance \cite{H.W.}. This amounts in the continuum to fixing the
gauge as the geodesic gauge.
A similar approach developed from the Mandelstam path space formulation
\cite{Mandelstam} has been recently given \cite{T.W.,Teitel}.
The aim of the present paper is to compute the propagators in such a gauge.
As we shall see
such a problem is not at all trivial because a simple minded computation of
 the  propagator gives meaningless, i.e. infinite results.\par

\indent It is well known that gravity can be formulated in the second or first
order approach and we shall derive the propagators in both formulations.
Formally in the second order approach one starts from the  functional
integral
\begin{equation}
\int e^{\int\sqrt{g} R(x) d^N x}\mu[g_{\mu\nu}]{\cal D}[g_{\mu\nu}]
\end{equation}
where the fundamental variable is the metric tensor $g_{\mu\nu}$ and $
\mu[g_{\mu\nu}]$
is the integration
measure which we do not need to specify here \cite{misura}.
We shall show in section 2 that
the radial gauge condition in the second order formalism can be written as
$x^\mu g_{\mu\nu}=x^\mu\eta_{\mu\nu}$. Thus the sharp gauge fixing will be
given by
$\delta(x^\mu(g_{\mu\nu}-\eta_{\mu\nu}))$ which is to be accompanied by the
relative Faddeev-Popov terms by adding to the lagrangian the expression
\begin{equation}
\bar{c}^\nu(x)x^\mu(\nabla_\mu c_\nu+\nabla_\nu c_\mu)(x).
\end{equation}

Note that contrary to what happens in the usual gauge theories
formulated
in the radial gauge, the ghosts do not decouple. In the first order
formalism on the other hand, one considers as fundamental variables the
vierbeins
$\tau^a_\mu$ and the connections $\Gamma^{ab}_\mu$. The functional integral
becomes
\begin{equation}
\int e^{-{1\over 2\kappa^2}\int(d\Gamma+\Gamma\wedge\Gamma)^{ab}
\wedge e^c \dots \epsilon_{abc\dots}}\mu[e^a_\mu]{\cal
D}[\Gamma^{ab}_\mu]
{\cal D}
[e^a_\mu].
\end{equation}
The natural way to introduce the radial gauge fixing without breaking the
symmetry
between the connections and the vierbeins in obtained by the
$N + N(N-1)/2$  gauge conditions $\delta(x^\mu \Gamma^{ab}_\mu)~
\delta(x^\mu (e^a_\mu-\delta^a_\mu))$
and supplying the relative Faddeev-Popov terms. As it happens in
Yang-Mills theory,
the ghost  associated to the local Lorentz symmetry formally decouple,
while those
related to the diffeomorphisms survive as in the second order  case.\par

\indent The technique for deriving the propagator in the radial gauge in a
generalization of
the gauge projection method developed in \cite{M.S.} for the usual gauge
theory case. We
recall that given a generic field one can define several projections to the
radial
 gauge that differ by the behavior of the projected field at the origin
and at
infinity. One has to keep in mind that all the problem is to give a
solution to
the radial Green's equation, i.e. the equation arising when one performs in
the
gravitational field lagrangian in presence of sources, an arbitrary
variation of
the gravitational field subject to the radial gauge conditions; such a
solution has to be symmetric in the field argument and radial. In section 2
 we give a
general construction of the propagator in any  ``sharp gauge'' of which the
radial
gauge is an example,  starting from the propagator in a generic gauge (e.g.
a gauge
of the Feynman-Landau type). Care however has to be exerted in the projection
process,
 because due to the singular nature of the correlation function at short
distances,
 spurious infinite solutions of the homogeneous equation can develop. In
fact
the
most natural propagator in the radial gauge would be the correlator
\begin{equation}
\langle P^0[h]_{\mu\nu}(x) P^0[h]_{\rho\sigma}(y)\rangle
\end{equation}
that corresponds to the selection, in the projection procedure, of the
field which
is regular at the origin. In practice such  a field is given by a proper
average
of the Feynman field along the geodesics joining the points $x$ and $y$ to
the
origin
 and due to the singular nature of the Riemann correlators at short
distances,
$d^{-N-2}$, it makes the written correlation function divergent in all
dimensions.\par

\indent Similarly the radial projected field that is regular at infinity,
$P^\infty
[h]_{\mu\nu}$, leads to
a propagator divergent in all $N\leq 4$, for infrared problems. On the
other
hand
it is shown in section 2 that one can consider a projected equation that
treats
the
origin and the infinity in symmetrical way. A singular behavior of the
field
$g_{\mu\nu}$
at the origin and/or at infinity is unavoidable as can be seen by
considering a
special family of Wilson loops described in section 4. A  formal solution to
the radial
$P^S$ projected Green's function is given by
\begin{equation}
\langle P^S[h]_{\mu\nu}(x) P^S[h]_{\rho\sigma}(y)\rangle
\end{equation}
where $P^S=(P^0+P^\infty)/2$,
which however still contains an infinite gauge term, solution of the
homogeneous
equation that has to be subtracted away. The result of such a subtraction
procedure
gives for the solution of the radial $P^S$ projected Green's equation
\begin{equation}
{1\over 2}\langle P^0[h]_{\mu\nu}(x) P^\infty[h]_{\rho\sigma}(y)\rangle+
{1\over 2}\langle P^\infty[h]_{\mu\nu}(x) P^0[h]_{\rho\sigma}(y)\rangle
\end{equation}
which is symmetric in the field arguments and finite for all $N>2$ and
explicitly
solves the problem of finding the propagator in the radial gauge.
In addition it is immediately verified that the given propagator satisfies
the radiality condition $x^\mu G_{\mu\nu,\rho\sigma}(x,y)=0$. A similar
procedure
works
successfully also in the first order  formalism.
{}From the technical viewpoint the transition from the propagator in a gauge
of
the
Feynman-Landau type to the radial gauge propagator in obtained by
expressing the
radial
fields in terms of the linearized Riemann and torsion tensors which are
invariant
under linearized gauge transformations; analytically the radial propagators
are
naturally  expressed in terms of
hypergeometric functions. In three dimensions the propagators acquire a
particularly
simple structure: in the first order approach the correlators of two
connections
vanish
identically while the correlator of two dreibeins or of a dreibein and a
connection
reduce to collinear contributions. In the second order formalism the
metric-metric
correlation function is also zero except for collinear contributions. This
is
clearly
related to the absence of propagating gravitons in three dimensions and to
the
physical
nature of the considered gauge.\par

\indent The paper is structured as follows: in section 2 we develop the general
procedure
for
deriving the propagators in sharp gauges from the ones in a generic gauge.
In
section  3
we formulate the radial condition in the second order formalism and relate
it to
the
 usual definition of the Riemann-Eisenhart coordinates; special attention
is
paid to
the allowed singularities of the metric at the origin and behavior at
infinity.
Then
we write
down the radial projectors in terms of the linearized Riemann and torsion
tensors and
finally we derive the propagator. In section 4 we repeat the same procedure in
first
order case, underlining the most important differences between the two
approaches.
Section 5 is devoted to the conclusions.  In Appendix A we show how to
express the
propagators
in terms of hypergeometric functions and work out explicitly the behavior
of the
propagators at the origin and infinity. Appendix B outlines the $\beta\neq
0$
(i.e.  non sharp) radial gauge in the second order  formalism.
\section{Propagators.}

In this section we shall develop a general procedure for the construction
of the propagators in a class of ``sharp'' gauge conditions starting
from the propagator given in a generic gauge. Subsection 2.A contains
 the general treatment; in subsections
2.B, 2.C, 2.D we display the concrete form
of the various operators in the case of Electrodynamics and linearized
Einstein theory in the second and first order formalism, respectively.

\subsection{General formalism.}

Let the equation of motion for a generic gauge field $A(x)$ be written in the
form
\begin{equation}
  K_x \, A(x) = J(x),
\label{dft}
\end{equation}
where $K$ is a linear, non-invertible, hermitean ``kinetic'' operator and
$J(x)$ is an external source coupled to $A$. The gauge transformations
of $A$ have the form
\begin{equation}
  A(x) \rightarrow A(x) + C_x \, f(x).
\label{njh}
\end{equation}
Here $C_x$ is another linear operator, which has the properties
\begin{equation}
  K_x \, C_x = 0 \qquad \mbox{(gauge invariance)}
\label{hre}
\end{equation}
and
\begin{equation}
  C^\dagger_x \, K_x = 0 \,\,{\rm which~~ implies}\, C^\dagger_x \, J(x) = 0
  \qquad \mbox{(``source conservation'')}.
\label{dou}
\end{equation}

We assume that we can  always add to the kinetic operator $K$ an
operator $K^{\cal F}$ that makes $K$ invertible, as it happens, for example,
in the Feynman gauge. We assume $K^{\cal F}$
to be of the form ${\cal F}^\dagger {\cal F}$, as deriving from a
quadratic gauge fixing $\int d x \, {\cal F}^2(A(x))$. ${\cal F}$ is meant
to be a linear operator on $A$. The propagator $G^{\cal F}$ corresponding to
this gauge fixing satisfies
\begin{equation}
  (K_x + K^{\cal F}_x) \, G^{\cal F}(x,y) = \delta(x-y)
\label{gre}
\end{equation}
and has the following property
\begin{equation}
  \int dy \, K^{\cal F}_x \, G^{\cal F}(x,y) \, J(y)=0
  \qquad \mbox{if} \ \ C^\dagger_y \, J(y) = 0.
\label{bsp}
\end{equation}
In other words, $K^{\cal F}_x$ vanishes when applied to the  fields
generated by physical sources. In fact, applying $C^\dagger$ to (\ref{gre})
we get
\begin{equation}
  C^\dagger_x \, K_x \, G^{\cal F}(x,y) +
  C^\dagger_x \, {\cal F}^\dagger_x \, {\cal F}_x \, G^{\cal F}(x,y)
  = C^\dagger_x \, \delta(x-y).
\end{equation}
But using (\ref{dou})
and integrating on a conserved source $J$ we obtain
\begin{equation}
  \int dy \, C^\dagger_x \, {\cal F}^\dagger_x \, {\cal F}_x \,
  G^{\cal F}(x,y) \, J(y) = 0.
\label{chy}
\end{equation}
We notice that ${\cal F} \, C$ is the kinetic ghost operator and as
such invertible. Then from (\ref{chy}) we get
\begin{equation}
  \int dy \, {\cal F}_x \, G^{\cal F}(x,y) \, J(y) = 0
\end{equation}
and multiplying by ${\cal F}^\dagger_x$ we finally prove (\ref{bsp}).


Let us now impose on $A$ a generic ``sharp'' gauge condition ${\cal G}$
\begin{equation}
  A^{\cal G}=\{A(x):\, {\cal G}(A(x))=0\}.
\label{bhe}
\end{equation}
${\cal G}$ is meant to be a linear function of $A$ and of its derivatives.
The field $A^{\cal G}(x)$ can be obtained from a generic field $A(x)$
through a (generally non-local) projector $P^{\cal G}$
\begin{equation}
  A^{\cal G}(x) = P^{\cal G}[A](x) = A(x) + C_x \, F^{\cal G} [A](x).
\label{puy}
\end{equation}
We require this projector to be insensitive to any ``previous gauge''
of the field, namely to satisfy
\begin{equation}
   P^{\cal G}[C_x \, f](x) = 0, \qquad \mbox{for any} \ f(x).
\label{cht}
\end{equation}
We consider now the adjoint projector ${P^{\cal G}}^\dagger$ that in general
does not coincide with $P^{\cal G}$; an exception is given by
Feynman-Landau type of gauges.
We shall now prove the following properties of the adjoint projector.

\medskip
\noindent {\it (1) ${P^{\cal G}}^\dagger$ produces conserved
sources} \cite{B.D.}.\par
\medskip
\noindent For, integrating (\ref{cht}) on a current $J(x)$ we have
\begin{equation}
  \int dx \, P^{\cal G}[C_x \, f](x) \, J(x) = 0;
\end{equation}
by definition of the adjoint projector, this means that
\begin{equation}
  \int dx \, f(x) \, C^\dagger_x \, {P^{\cal G}}^\dagger[J](x)= 0
\end{equation}
and due to the arbitrariness of $f(x)$
\begin{equation}
  C^\dagger_x {P^{\cal G}}^\dagger[J](x) = 0 \qquad \mbox{for any} \ J(x).
\label{xpf}
\end{equation}

\noindent {\it (2) ${P^{\cal G}}^\dagger$ leaves
a conserved source unchanged.}\par
\medskip
\noindent Let us suppose that $J$ is conserved. Using (\ref{puy})
we have for a generic $A$
\begin{equation}
  \int dx \, A(x) \, {P^{\cal G}}^\dagger [J](x) =
  \int dx \, A(x) \, J(x) + \int dx \, \{ C_x \, F^{\cal G}[A](x) \} \, J(x) .
\end{equation}
Integrating by parts the second term on the r.h.s. and using
 (\ref{dou}) and the arbitrariness of $A$ we have
\begin{equation}
  {P^{\cal G}}^\dagger[J](x) = J(x) \qquad \mbox{if} \ C^\dagger_x \, J(x) = 0.
\label{vls}
\end{equation}

\bigskip
The equation of motion obtained varying the action under the constraint
(\ref{bhe}) is
\begin{equation}
  {P^{\cal G}}^\dagger[K_x \, A^{\cal G}](x) = {P^{\cal G}}^\dagger[J](x).
\label{vro}
\end{equation}
{}From (\ref{dou}) and Property 2 we have
\begin{equation}
  K_x \, A^{\cal G}(x) = {P^{\cal G}}^\dagger[J](x),
\label{vpe}
\end{equation}
or for the propagator
\begin{equation}
  K_x \, G^{\cal G}(x,y) = {P^{\cal G}}^\dagger[\delta(x-y)],
\label{lpi}
\end{equation}
where the meaning of the r.h.s. is
\begin{equation}
  \int dy \, {P^{\cal G}}^\dagger[\delta(x-y)] \, J(y) =
  {P^{\cal G}}^\dagger[J](x).
\end{equation}
Next we show that a solution of (\ref{lpi}) is
\begin{equation}
  G^{\cal G}(x,y) = i\left< P^{\cal G}[A^{\cal F}](x) \,
  P^{\cal G}[A^{\cal F}](y) \right>_0,
\label{jhg}
\end{equation}
where $A^{\cal F}$ denotes the field in the original gauge.
Integrating on a source $J(y)$ we have
\begin{eqnarray}
  & & i\int dy \, K_x \left< P^{\cal G}[A^{\cal F}](x) \,
  P^{\cal G}[A^{\cal F}](y) \right>_0 \, J(y) = \nonumber \\
  & & = i\int dy \, K_x \left< P^{\cal G}[A^{\cal F}](x) \,
  A^{\cal F}(y) \right>_0 \, {P^{\cal G}}^\dagger[J](y) = \nonumber \\
  & & = i\int dy \, K_x \left< \{ A^{\cal F}(x) +
  C_x \, F^{\cal G}[A^{\cal F}](x) \} \,
  A^{\cal F}(y) \right>_0 \, {P^{\cal G}}^\dagger[J](y) = \nonumber \\
  & & = i\int dy \, K_x \left< A^{\cal F}(x) \,
  A^{\cal F}(y) \right>_0 \, {P^{\cal G}}^\dagger[J](y) = \nonumber \\
  & & = \int dy \, (K_x + K^{\cal F}_x) \, G^{\cal F}(x,y) \,
  {P^{\cal G}}^\dagger[J](y) -
  \int dy \, K^{\cal F}_x \, G^{\cal F}(x,y) \, {P^{\cal G}}^\dagger[J](y)
  = \nonumber \\
  & & = \int dy \, \delta(x-y) \, {P^{\cal G}}^\dagger[J](y).
\label{cmd}
\end{eqnarray}
In the last step we have used (\ref{bsp}) and Property 1.
Note however that $G^{\cal G}$ defined in
(\ref{jhg}) remains unchanged if we replace the original field with
any other gauge equivalent field.

Given two different projectors $P_1$ and $P_2$ that project on the same
gauge (which as a rule differ for different boundary conditions), one has
$P_1 \, P_2 = P_1$ and $P_2 \, P_1 = P_2$ due to eq.
(\ref{cht})\footnotemark. \footnotetext{The definition of $P^0$ and
$P^{\infty}$ is different from one
adopted in ref. \cite{M.S.}, and they obey a different algebra. The
definition adopted here privileges the invariance under gauge
transformations on the original fields while the ones given in \cite{M.S.}
privileges the radiality. These projectors give the same results when
applied to regular fields.}  Thus
also $P_{12}=\alpha P_1 + (1-\alpha) P_2$ is
a projector on the
considered gauge and one can write down the $P_{12}$-projected Green
function equation (we omit the suffix ${\cal G}$)
\begin{equation}
  K_x \, G(x,y) = P^\dagger_{12}[\delta(x-y)].
\label{xgh}
\end{equation}
It is immediate to verify that a solution of (\ref{xgh}) is also  given by
\begin{equation}
  \alpha \, \left< P_2[A](x) \, P_1[A](y) \right>_0 +
  (1-\alpha) \left< P_1[A](x) \, P_2[A](y) \right>_0.
\label{fgz}
\end{equation}
In fact
\begin{equation}
  K_x \, \alpha \, \left< P_2[A](x)\, P_1[A](y) \right>_0 =
  \alpha \, K_x \, \left< A(x) \, P_1[A](y) \right>_0 = \alpha \, P^\dagger_1
  [\delta(x-y)],
\end{equation}
repeating the same procedure of eq. (\ref{cmd}). Acting similarly with
the $(1-\alpha)$ term in (\ref{fgz}), we get (\ref{xgh}). We notice that
for $\alpha = \frac{1}{2}$, \  (\ref{fgz}) is symmetric in the exchange
of the field arguments.
In the following we write  the explicit form of the  operators appearing in
the Electrodynamics and linearized Einstein theory
\subsection{Electrodynamics and linearized Yang-Mills theory.}

This is the most simple case. We have the following identifications
\begin{eqnarray}
  & & A \rightarrow A_\mu; \\
  & & J \rightarrow J_\mu; \\
  & & C f \rightarrow \partial_\mu \, f; \\
  & & K \rightarrow \eta_{\mu \nu} \,
  \Box - \partial_\mu \partial_\nu; \\
  & & K^{\cal F} \rightarrow \partial_\mu \partial_\nu,
\end{eqnarray}
where $K^{\cal F}$ is the operator produced by the usual
Feynman gauge fixing $\frac{1}{2}(\partial^\mu A_\mu)^2$.

\subsection{Linearized Einstein gravity in the second-order formalism.}

In this case we have
\begin{eqnarray}
  & & A \rightarrow h_{\mu \nu}; \\
  & & J \rightarrow T_{\mu \nu}; \\
  & & C f \rightarrow \left( \delta_{\alpha \sigma}
  \partial_\rho + \delta_{\alpha \rho} \partial_\sigma \right)
  f_\alpha ; \\
  & & K \rightarrow
  K_{\mu \nu \rho \sigma} = \frac{1}{4} [2 \, \eta_{\rho \sigma} \,
  \partial_\mu \partial_\nu +
  2 \, \eta_{\mu \nu} \, \partial_\rho \partial_\sigma + \nonumber \\
  \label{mkw}
  & & \qquad \qquad \qquad -
  ( \eta_{\mu \rho} \, \partial_\nu \partial_\sigma +
  \eta_{\nu \rho} \, \partial_\mu \partial_\sigma
  + \eta_{\mu \sigma} \, \partial_\nu \partial_\rho +
  \eta_{\nu \sigma} \, \partial_\mu \partial_\rho ) + \nonumber \\
  & & \qquad \qquad \qquad + (\eta_{\mu \rho} \, \eta_{\nu \sigma}
  + \eta_{\mu \sigma} \, \eta_{\nu \rho}
  - 2 \, \eta_{\mu \nu} \, \eta_{\rho \sigma}) \, \Box ] ; \\
  & & K^{\cal F} \rightarrow
  K^{\cal F}_{\mu \nu \rho \sigma} =
  \frac{1}{4} [ - ( 2 \, \eta_{\rho \sigma} \,
  \partial_\mu \partial_\nu +
  2 \, \eta_{\mu \nu} \, \partial_\rho \partial_\sigma ) +
  \nonumber \\
  & & \qquad \qquad \qquad
  + ( \eta_{\mu \rho} \, \partial_\nu \partial_\sigma +
  \eta_{\nu \rho} \, \partial_\mu \partial_\sigma
  + \eta_{\mu \sigma} \, \partial_\nu \partial_\rho +
  \eta_{\nu \sigma} \, \partial_\mu \partial_\rho) ] .
\end{eqnarray}
Here $K^{\cal F}$ is the operator produced by the harmonic gauge fixing
\begin{equation}
  \frac{1}{2} \left( \partial^\mu h_{\mu \nu} -
  \frac{1}{2} \partial^\nu h^\mu_\mu \right)^2.
\label{cre}
\end{equation}

\subsection{Linearized Einstein gravity in the first-order formalism.}

The quadratic part of the lagrangian has the form
\begin{equation}
  L^{(2)} = - (
  \delta^{\mu \nu \gamma}_{abc} \, \partial_\mu \, \Gamma^{ab}_\nu
  \, \tau^c_\gamma + \delta^{\mu \nu}_{ab}
  \, \Gamma^a_{c \mu} \, \Gamma^{cb}_\nu
  + T^\mu_a \, \tau^a_\mu + \Sigma^\mu_{ab} \, \Gamma^{ab}_\mu ) ,
\end{equation}
where
\begin{equation}
  \tau^a_\mu = e^a_\mu - \delta^a_\mu
\end{equation}
and $T^\mu_a$ and $\Sigma^\mu_{ab}$ are the energy-momentum source and
the spin-torsion source, respectively. The gauge transformations have the form
\begin{equation}
  Cf \rightarrow \left(
  \begin{array}{cc}
    0 & \ \ \delta^{ab}_{cd} \, \partial_\mu\\
    & \\
    \partial_\mu & \ \ - \eta_{\mu b} \, \delta^{ab}_{cd}
  \end{array} \right)
  \left(
  \begin{array}{c}
    \Lambda^a \\
    \\
    \theta^{cd}
  \end{array} \right).
\end{equation}
The field equations are given by
\begin{equation}
  K \, A = J \rightarrow \left(
  \begin{array}{cc}
    \frac{1}{2} (\eta_{c \sigma} \delta^{\nu \sigma \mu}_{dab}
    - \eta_{d \sigma} \delta^{\nu \sigma \mu}_{cab}) & \ \
    - \delta^{\lambda \gamma \mu}_{rab} \, \partial_\lambda  \\
    & \\
    \delta^{\mu \lambda \nu}_{acd} \, \partial_\lambda & 0
  \end{array} \right)
  \left(
  \begin{array}{c}
    \Gamma^{cd}_{\nu} \\
    \\
    \tau^r_\gamma
  \end{array} \right) =
  \left(
  \begin{array}{c}
    \Sigma^\mu_{ab} \\
    \\
    T^\mu_a
  \end{array} \right)
\end{equation}
and the gauge-fixing term has the form
\begin{equation}
  K^{\cal F} \, A \rightarrow
  \left(
  \begin{array}{cc}
  0 & \ \ 0 \\
  & \\
  0 & \ \ 4K^{\cal F}_{a \mu r \nu} \, \eta^{\nu \gamma}
  + 2\beta \, (\eta_{ar} \eta^{\mu \gamma} - \delta^\mu_r \delta^\gamma_a)
  \end{array} \right)
  \left(
  \begin{array}{c}
  \Gamma^{cd}_{\nu} \\
  \\
  \tau^r_\gamma
  \end{array} \right),
\end{equation}
and is produced by the harmonic gauge fixing (\ref{cre}) with
$h_{\mu \nu}=\tau^a_\mu \eta_{a \nu} + \tau^a_\nu \eta_{a \mu}$,
to which the symmetric gauge fixing $ \frac{\beta}{2}
(\tau^a_\mu \eta_{a \nu} - \tau^a_\nu \eta_{a \mu})^2$ has been
added.

\section{Second order formalism.}

\subsection{Radial gauge.}

We first give a discussion at the classical level.
In the second order formalism we define the radial gauge  through
the  condition
\begin{equation}
  \label{radsec}
  \xi^\mu g_{\mu\nu}(\xi)=\xi^\mu\eta_{\mu\nu}
\end{equation}
($\eta_{\mu\nu}={\rm diag}(1,-1,-1,\dots)$, and $N=$space-time dimensions).
Taking the derivative of (\ref{radsec}) with respect to $\xi^\lambda$ we
have
\begin{equation}
  \label{dereq1}
  g_{\lambda\mu}(\xi)+\xi^\mu \partial_\lambda
  g_{\mu\nu}(\xi)=\eta_{\lambda\nu}.
\end{equation}
Under regularity hypothesis on $g_{\mu\nu}(\xi)$ we obtain
\begin{equation}
\label{g(0)}
  g_{\lambda\nu}(0)=\eta_{\lambda\nu}.
\end{equation}
Contracting the Levi-Civita connection on $\xi^\alpha$ and $\xi^\beta$ we get,
using (\ref{radsec}) and (\ref{dereq1})
\begin{eqnarray}
  \label{LevCiv}
  & & \xi^a\xi^\beta\Gamma_{\mu,\alpha\beta}(\xi) =
  \frac{1}{2}\xi^\alpha\xi^\beta
  [ \partial_\beta g_{\mu\alpha}(\xi)+\partial_\alpha
  g_{\mu\beta}(\xi)-\partial_\mu g_{\alpha\beta}(\xi)]=\nonumber\\
  & & \qquad = \frac{1}{2} \xi^\beta[\eta_{\beta\mu} -g_{\beta\mu}(\xi)]
  + \frac{1}{2} \xi^\alpha[\eta_{\alpha\mu}-g_{\alpha\mu}(\xi)]
  - \frac{1}{2} \xi^\beta[\eta_{\beta\mu}-g_{\beta\mu}(\xi)]=0.
\end{eqnarray}
Thus coordinates satisfying (\ref{radsec}) are Riemann's normal
coordinates, in the sense that the lines $\xi^\mu = \lambda n^\mu$
are autoparallel lines \cite{H.E.}. Furthermore (\ref{LevCiv})
combined with (\ref{radsec}) tells us that $\xi^\mu = \lambda n^\mu$
are also geodesic in the sense that they are extrema of the distance
between two events. In fact from (\ref{radsec}) we have that
\begin{equation}
  (ds)^2 = d\xi^\mu \, g_{\mu \nu}(\xi) \, d\xi^\nu =
  (d\lambda)^2 \, n^\mu \, g_{\mu \nu}(\xi) \, n^\nu =
  (d\lambda)^2 \, n^\mu \, g_{\mu \nu}(\xi)\vert_{\xi=0} \, n^\nu
\end{equation}
and thus $\xi^\mu(s)$ satisfies the equation
\begin{equation}
\frac{d^2\xi^\mu}{d
s^2}+\Gamma^{\mu}_{\alpha\beta}(\xi)\frac{d\xi^\alpha}{d s} \frac{d
\xi^\beta}{d s}=0.
\end{equation}
\noindent The usual definition of the geodesic gauge is \cite{M.T.W.}
\begin{equation}
  \label{radsec1}
  \xi^\alpha \xi^\beta\Gamma_{\mu,\alpha\beta}(\xi)=0,
\end{equation}
as given by eq. (\ref{LevCiv}).
We want to prove that  eq. (\ref{radsec}) follows from (\ref{radsec1})
(apart from a global linear transformation). In fact from (\ref{radsec1})
we have
\begin{eqnarray}
  \label{Delta}
  & & \frac{1}{2}\xi^\alpha\xi^\beta
  [\partial_\beta g_{\mu\alpha}(\xi)+\partial_\alpha
  g_{\mu\beta}(\xi)-\partial_\mu g_{\alpha\beta}(\xi)] =  \nonumber\\
  & & \qquad = \xi^\alpha \partial_\alpha[\xi^\beta g_{\beta\mu}(\xi)]
  -\frac{1}{2}\partial_\mu [\xi^\alpha\xi^\beta
  g_{\alpha\beta}(\xi)]=0.
\end{eqnarray}
Multiplying (\ref{Delta}) by $\xi^\mu$ we obtain
\begin{equation}
  \frac{1}{2}\xi^a\partial_\alpha[\xi^\beta\xi^\mu
  g_{\beta\mu}(\xi)]-\xi^\beta\xi^\mu g_{\beta\mu}(\xi)=0,
\end{equation}
i.e. $\xi^\mu\xi^\beta g_{\mu\beta}(\xi)$ is a homogeneous
function of degree 2 in $\xi$.
In the following (see subsection 3.C) we shall consider metrics $g_{\mu\nu}$
which a priori  are not regular at the origin but such that $\xi^\mu
g_{\mu\beta}(\xi) $ is a regular ($C^2$) function  in a domain containing
the origin and which vanish for $\xi^\mu=0$. Under such assumption we
have
\begin{equation}
  \xi^\beta \xi^\mu g_{\beta \mu}(\xi) = c_{\mu \beta} \xi^\beta \xi^\mu .
\end{equation}
In fact taking the second derivative we have
\begin{equation}
  \partial_\rho \partial_\lambda
  [ \xi^\beta \xi^\mu g_{\beta \mu}(\xi) ] =
  H^0_{\rho \lambda}(\xi) = c_{\rho \lambda} ,
\end{equation}
because the only continuous homogeneous function of degree $0$ is
the constant.

\noindent Substituting this relation into (\ref{Delta}) we have
\begin{equation}
  \xi^a\partial_a[\xi^\beta g_{\beta\mu}(\xi)
  - \xi^\beta c_{\beta\mu}] = 0,
\end{equation}
i.e. $\xi^\beta g_{\beta\mu}(\xi)-\xi^\beta c_{\beta\mu}$ must be an
homogeneous
function of degree 0, that has to vanish for $\xi=0$ and is thus
identically zero
\begin{equation}
\xi^\beta g_{\beta\mu}(\xi)=\xi^\beta c_{\beta\mu}.
\end{equation}
In conclusion we have found that (\ref{radsec1}) implies (\ref{radsec})
up to a global linear transformation of the coordinates.
In the following we shall adopt (\ref{radsec}) as the definition of the radial
gauge in the second order formalism.

\subsection{Radial Projectors.}
In this section we shall derive, given the linearized Riemann tensor, radial
fields $h^{0}_{\mu\nu}(x)$ and $h^\infty_{\mu\nu}(x)$ which generate such a
Riemann tensor. $h^{0}_{\mu\nu}(x)$ and $h^\infty_{\mu\nu}(x)$ differ for
their regularity properties at the origin and infinity and they will be the
ingredients out of which we construct the  Green's function.
We now want to express the radial metric
$g_{\mu\nu}(x)=\eta_{\mu \nu}+h_{\mu \nu}(x)$ in
terms of the Riemann tensor, at the linearized level.
Let us start from the expression of the linearized Riemann tensor
\begin{equation}
  \label{Riemann}
  R^{L}_{\mu\nu,\alpha\beta}(x) = \frac{1}{2}
  [\partial_\alpha\partial_\nu h_{\mu\beta}(x)
  - \partial_\alpha\partial_\mu h_{\nu\beta}(x)
  - \partial_\beta\partial_\nu h_{\mu\alpha}(x)
  + \partial_\beta\partial_\mu h_{\alpha\nu}(x)].
\end{equation}
 $R^{L}_{\mu\nu,\alpha\beta}(x)$ is an invariant under the
linearized reparametrization transformations. Using the radial condition
(\ref{radsec}) in the form
\begin{equation}
  \label{radsec2}
  {\cal G}(h)_\nu(x) = x^\mu h_{\mu\nu}(x) = 0
\end{equation}
we can rewrite the contraction of (\ref{Riemann}) with $x^\mu \, x^\alpha$
as follows
\begin{equation}
  2 \, x^\mu x^\alpha R^{L}_{\mu\nu,\alpha\beta}(x) =
  - x^\mu \partial_\mu [x^\alpha \partial_\alpha h_{\nu\beta}(x)] -
  x^\alpha \partial_\alpha h_{\nu\beta}(x).
\end{equation}
Applying the usual change of variables $x\rightarrow\lambda x$ we get
\begin{eqnarray}
  \label{Delta1}
  & & -2 \, \lambda^2  x^\mu  x^\alpha
  R^{L}_{\mu \nu,\alpha \beta}(\lambda  x) = \nonumber \\
  & & \qquad  =\lambda x^\mu \partial_\mu [\lambda  x^\alpha
  \partial_\alpha h_{\nu \beta}(\lambda  x)] +
  \lambda  x^\alpha \partial_\alpha h_{\nu \beta}(\lambda  x)=
  \frac{d}{d\lambda} \left[ \lambda^2  x^\alpha \partial_\alpha
  h_{\nu\beta}(\lambda x) \right].
\end{eqnarray}
  Integrating (\ref{Delta1}) in $\lambda$ from $0$ to $1$ we obtain
\begin{equation}
  \label{Delta2}
   x^\alpha \partial_\alpha h_{\nu\beta}( x) =
  - 2 \,  x^\alpha  x^\mu \int^1_0 d\lambda \, \lambda^2
  R^{L}_{\mu\nu,\alpha\beta}(\lambda x),
\end{equation}
provided the metric is such that $|R^L( x)| \le | x|^{-3+\varepsilon}$
as $ x \to 0$. With $\vert x \vert^\alpha$ we understand here a quantity that
for $x^\mu=\lambda x^\mu$ behaves like $\lambda^\alpha $ for $\lambda\to 0$.
We notice again that
\begin{equation}
  \tau  x^\alpha \partial_\alpha h_{\nu \beta}(\tau  x) =
  \tau \frac{d}{d\tau} \left[ h_{\nu \beta}(\tau  x) \right]
\end{equation}
and thus we have
\begin{eqnarray}
  \label{Delta3}
  h^0_{\nu\beta}( x) & = & -2 \,  x^\alpha  x^\mu \int^1_0
  d\tau \, \tau \int^1_0 d\lambda \, \lambda^2 \,
  R^{L}_{\mu \nu,\alpha \beta}(\lambda \tau  x) = \nonumber \\
  & = & -2 \,  x^\alpha  x^\mu \int^1_0 d\lambda \, \lambda (1-\lambda)
  \, R^{L}_{\mu \nu,\alpha \beta}(\lambda  x) = \nonumber \\
  & = & P^0[h]_{\nu \beta}(x),
\end{eqnarray}
provided $\vert R^L(x)\vert< \vert x\vert^{-2+\varepsilon}$ for $x\to 0$.
The superscript ``0'' on $h_{\nu\beta}( x)$ specifies the integration limits.
Another possibility to solve equation (\ref{Delta1}) is to integrate twice
from $1$ to $\infty$, provided the metric is such that
$|R^L( x)| \le | x|^{-3-\varepsilon}$ as $ x \to \infty$. Then we have
\begin{eqnarray}
\label{Delta4}
  h^{\infty}_{\nu\beta}( x) & = & - 2 \,  x^\alpha  x^\mu
  \int^\infty_1 d\tau \, \tau \int^\infty_1 d\lambda \, \lambda^2 \,
  R^{L}_{\mu\nu,\alpha\beta}(\lambda\tau x) = \nonumber \\
  & = & - 2 \,  x^\alpha  x^\mu \int^\infty_1 d\lambda \, \lambda
  (\lambda-1) \, R^{L}_{\mu\nu,\alpha\beta}(\lambda x) = \nonumber \\
  & = & P^\infty[h]_{\nu \beta}(x).
\end{eqnarray}
Eqs. (\ref{Delta3}), (\ref{Delta4}) define projectors from an arbitrary
gauge to the radial gauge. The projector nature of $P^0$ and $P^\infty$
is proven by showing that e.g. $h$ and $P^0[h]$ give rise to the same
Riemann tensor. In fact by using the linearized Bianchi identities
\begin{equation}
  \partial_\lambda R_{\mu \nu , \alpha \beta}(x) +
  \partial_\mu R_{\nu \lambda , \alpha \beta}(x) +
  \partial_\nu R_{\lambda \mu , \alpha \beta}(x) = 0
\end{equation}
and the familiar symmetry properties of the Riemann tensor, one reduces
${R^L}^0_{\mu \nu , \alpha \beta}(x)$  to
\begin{equation}
  {R^L}^0_{\mu \nu , \alpha \beta}(x) = \int_0^1 d\lambda \,
  \frac{d}{d\lambda} [\lambda^2 \,
  R^L_{\mu \nu , \alpha \beta}(\lambda x)].
\label{cxz}
\end{equation}
Whenever $P^0$ is defined on $h$ we have from (\ref{cxz})
\begin{equation}
   {R^L}^0_{\mu \nu , \alpha \beta}(x) = R^L_{\mu \nu , \alpha \beta}(x).
\end{equation}
Furthermore from the gauge invariance of $R^L_{\mu \nu , \alpha \beta}(x)$
appearing in the definitions (\ref{Delta3}), (\ref{Delta4}) we have the
properties
\begin{equation}
  P^0 \, P^\infty = P^0, \qquad P^\infty \, P^0 = P^\infty.
\label{fut}
\end{equation}

The propagator
\begin{equation}
  \left< P^0[h]_{\mu \nu}(x) \, P^0[h]_{\rho \sigma}(y) \right>_0
\label{sog}
\end{equation}
computed by means of (\ref{Delta3}) is divergent in all dimensions due
to the $(x-y)^{-N-2}$ behavior of the correlation of two Riemann's tensors
at short distances. On the other hand
\begin{equation}
  \left< P^\infty[h]_{\mu \nu}(x) \, P^\infty[h]_{\rho \sigma}(y)
  \right>_0
\end{equation}
diverges for $N\le 4$ due to the  infrared behavior of the correlator
of two Riemann's tensors. For this reason we shall adopt
a different gauge projector given by
\begin{equation}
  P^S = \frac{1}{2} (P^0 + P^\infty)
\end{equation}
whose projector nature is immediately seen from (\ref{fut}). We elucidate
the nature of the boundary condition at $0$ and $\infty$ of the projector
$P^S$ at the end of section 3.C.

As we have shown in section 2 the solution of the projected Green's
function equation is given by
\begin{equation}
  \frac{1}{2} \left< P^0[h]_{\mu \nu}(x) \, P^\infty [h]_{\rho \sigma}(y)
  \right>_0 +
  \frac{1}{2} \left< P^\infty[h]_{\mu \nu}(x) \, P^0 [h]_{\rho \sigma}(y)
  \right>_0
\end{equation}
which as we shall show in the next section is convergent for $N>2$.

In order to write the source term in the projected Green's function
equation we must compute the adjoint of $P^S$, i.e. the adjoints of
$P^0$ and $P^\infty$. We have
\begin{eqnarray}
  & & {P^0}^\dagger[T]^{\mu \nu}(x) = T^{\mu \nu}(x) +
  \int_1^\infty d\tau \, \tau^N [x^\nu \partial_\rho T^{\mu \rho}(\tau x)
  + x^\mu \partial_\rho T^{\nu \rho}(\tau x)] + \nonumber \\
  & & \qquad \qquad \qquad + x^\mu x^\nu \int_1^\infty d\tau \, \tau^N (\tau-1)
  \partial_\alpha \partial_\beta T^{\alpha \beta}(\tau x), \nonumber \\
  & & \nonumber \\
  & & {P^\infty}^\dagger[T]^{\mu \nu}(x) = T^{\mu \nu}(x) -
  \int_0^1 d\tau \, \tau^N [x^\nu \partial_\rho T^{\mu \rho}(\tau x)
  +x^\mu \partial_\rho T^{\nu \rho}(\tau x)] + \nonumber \\
  & & \qquad \qquad \qquad - x^\mu x^\nu \int_0^1 d\tau \, \tau^N (\tau-1)
  \partial_\alpha \partial_\beta T^{\alpha \beta}(\tau x)
\end{eqnarray}
and
\begin{equation}
  {P^S}^\dagger = \frac{1}{2} ({P^0}^\dagger + {P^\infty}^\dagger).
\end{equation}

\subsection{Radial propagators.}

We shall now derive the propagators using the general procedure explained
in section 2. The kinetic operator $K$ is given by formula (\ref{mkw}).
The equation for the propagator is
\begin{equation}
  (K_x)_{\mu \nu}^{\rho \sigma} \, G_{\rho \sigma \alpha \beta}(x,y) =
  P^\dagger[\Delta_{\mu \nu \alpha \beta}(x-y)],
\end{equation}
where
\begin{equation}
  \Delta_{\mu \nu \alpha \beta}(x-y) =
  \frac{1}{2} \, (\eta_{\mu \alpha} \, \eta_{\nu \beta}
  + \eta_{\mu \beta} \, \eta_{\nu \alpha}) \, \delta^N(x-y).
\label{qpf}
\end{equation}

As a preliminary step we must compute the correlator of two Riemann
tensors; this is most simply obtained by using the harmonic gauge.
We find
\begin{eqnarray}
\label{RR2}
  & & \left< R_{\mu \nu , \alpha \beta}(x) \, R_{\rho \sigma , \lambda
  \gamma}(y) \right>_0 =
  \frac{1}{4} \delta^{\mu' \nu'}_{\mu \nu} \eta_{\nu' \nu''} \, (
  \delta^{\rho' \nu''}_{\rho \nu} \delta^{\alpha' \beta'}_{\alpha \beta}
  \delta^{\lambda' \beta''}_{\lambda \gamma}
  \eta_{\beta' \beta''}+ \nonumber \\
  & &+ \, \delta^{\lambda' \nu''}_{\lambda \gamma}
  \delta^{\alpha' \beta'}_{\alpha \beta} \delta^{\rho' \beta''}_{\rho \sigma}
  \eta_{\beta' \beta''} -
  \mbox{$\frac{2}{N-2}$} \,
  \delta^{\alpha' \nu''}_{\alpha \beta} \delta^{\rho' \sigma'}_{\rho \sigma}
  \delta^{\lambda' \sigma''}_{\lambda \gamma}
  \eta_{\sigma' \sigma''} ) \
  \partial_{\mu'} \partial_{\alpha'} \partial_{\rho'} \partial_{\lambda'}
  D(x-y)
\label{mpj}
\end{eqnarray}
being $D(x-y)$ the usual Feynman propagator
\begin{equation}
D(x)=\frac{\Gamma (N/2-1)}{4 \pi^{\frac{N}{2}}}\frac{i}{(x^2- i \varepsilon
 )^{N/2-1}}.
\end{equation}
If we choose $P=P^0$ and
substitute in (\ref{jhg}) eq. (\ref{mpj}) we find that the propagator can be
expressed as a linear combination of derivatives of integrals of the type
\begin{equation}
 F^0_{\alpha\beta}(x,y)= \int_0^1 d\lambda \, (1-\lambda) \int_0^1 d\tau \,
  (1-\tau) \partial_\alpha \partial_\beta D(\lambda x - \tau y).
\label{fsd}
\end{equation}
With the method explained below one sees that such integral diverges
in any dimension due to the bad ultraviolet behavior of
$\partial_\alpha \partial_\beta D(\lambda x - \tau y)$. Similarly
\begin{equation}
  G^\infty_{\mu \nu \rho \sigma}(x,y) =
  \left< P^\infty[h]_{\mu \nu}(x) \, P^\infty[h]_{\rho \sigma}(y)
  \right>_0
\end{equation}
computed along the same lines is expressed in terms of derivatives of
\begin{equation}
  F^\infty_{\alpha\beta}(x,y)=\int_1^\infty d\lambda \, (1-\lambda)
\int_1^\infty  d\tau \,
  (1-\tau) \partial_\alpha \partial_\beta D(\lambda x - \tau y)
\label{gff}
\end{equation}
and as shown below it diverges for $N \leq 4$.
On the other hand the correlator
\begin{equation}
  \left< P^0[h]_{\mu \nu}(x) \, P^\infty[h]_{\rho \sigma}(y)
  \right>_0
\label{fuy}
\end{equation}
used to construct the solution of the $P^S$ projected Green's equation
converges for $N > 2$.

We prove now the above stated results: the correlator (\ref{fuy}) can
be expressed as linear combination of derivatives of the double integral
\begin{equation}
\label{FS2}
  F^S_{\alpha \beta}(x,y) =
  \int_0^1 d\lambda \, (1-\lambda) \int_1^\infty d\tau \,
  (\tau - 1) \partial_\alpha \partial_\beta D(\lambda x - \tau y)
\label{dkc}
\end{equation}
where
\begin{equation}
  \partial_\alpha \partial_\beta D(x) =
  -(N-2) \, \frac{\eta_{\alpha \beta}}{\ (x^2)^{N/2}} +
  N(N-2) \, \frac{x_\alpha x_\beta}{\ \ (x^2)^{N/2+1}}.
\label{vgr}
\end{equation}
The first term of (\ref{vgr}) substituted in (\ref{dkc}) gives
\begin{equation}
  \int_0^1 d\lambda \, \lambda^{-N+1}(1-\lambda) \int_0^\lambda
  d\rho \, \left( \frac{\lambda}{\rho} - 1 \right)
  \frac{\rho^{N-2}}{(\rho x - y)^N}
\end{equation}
and converges for $N>2$. Similarly one proves the convergence for
$N>2$ of the second term of (\ref{vgr}) substituted in (\ref{dkc}).
Thus the solution of the $P^S$ projected Green's equation
\begin{equation}
  (K_x)_{\mu \nu}^{\rho \sigma} \, G^S_{\rho \sigma \alpha \beta}(x,y) =
  {P^S}^\dagger[\Delta_{\mu \nu \alpha \beta}(x-y)]
\end{equation}
given by
\begin{equation}
  -iG^S_{\mu \nu \rho \sigma}(x,y) =
  \frac{1}{2} \left< P^0[h]_{\mu \nu}(x) \, P^\infty[h]_{\rho \sigma}(y)
  \right>_0 +
  \frac{1}{2} \left< P^\infty[h]_{\mu \nu}(x) \, P^0[h]_{\rho \sigma}(y)
  \right>_0
\end{equation}
converges for all $N>2$.

With regard to the $G^\infty(x,y)$ given by (\ref{gff}) one reaches
for the analogue $F^\infty_{\alpha \beta}(x,y)$ of
$F^S_{\alpha \beta}(x,y)$ the integral
\begin{equation}
  (N-2)\int_1^\infty d\lambda \, \lambda^{-N+1}(1-\lambda) \int_0^\lambda
  d\rho \, \left( \frac{\lambda}{\rho} - 1 \right)
  \frac{\rho^{N-2}}{(\rho x - y)^N}\left ( -\eta_{\alpha\beta}+N
\frac{x_\alpha x_\beta}{(\rho x-y)^2}\right )
\end{equation}
which converges only for $N>4$. Finally the
function $F^0_{\alpha \beta}(x,y)$ given by
\begin{equation}
  F^0_{\alpha \beta}(x,y) =
  \int_0^1 d\lambda \, (1-\lambda) \int_0^1 d\tau \,
  (1-\tau) \partial_\alpha \partial_\beta D(\lambda x - \tau y)
\end{equation}
leads to integrals of the form
\begin{equation}
  \int_0^1 d\lambda \, (1-\lambda) \, \lambda^{-N+1}
  \int_\lambda^\infty d\rho \, \left( \frac{\lambda}{\rho} - 1 \right)
  \frac{\rho^{N-2}}{(\rho x - y)^N}
\end{equation}
which diverge for all $N \geq 2$.

In conclusion, we produced a solution of the radial Green's function
equation that is radial, symmetric and finite for $N>2$. Such a
solution treats the infinity and the origin in symmetrical way in the
sense that the field
\begin{equation}
  h^S_{\mu \nu}(x) = \int G^S_{\mu \nu \rho \sigma}(x,y) \,
  t^{\rho \sigma}(y) \, d^Ny ,
\end{equation}
computed on a physical (i.e. conserved) source $t^{\rho \sigma}$,
behaves like
\begin{equation}
  h^S(x) = K^0(x) + \frac{H^0(x)}{r} + O(r^\varepsilon)
  \qquad \mbox{for} \ \ r \rightarrow 0
\label{lsu}
\end{equation}
and
\begin{equation}
  h^S(x) = - K^0(x) - \frac{H^0(x)}{r} + O(r^{-1-\varepsilon})
  \qquad \mbox{for} \ \ r \rightarrow \infty ,
\label{kdj}
\end{equation}
where $K^0$ and $H^0$ are homogeneous functions of $x$ of degree 0,
and $r=|x|$.
In fact, given a conserved source $t^{\rho \sigma}(y)$ we have
\begin{eqnarray}
  & & -i \int G^S_{\mu \nu \rho \sigma}(x,y) \,
  t^{\rho \sigma}(y) \, d^Ny  =  \frac{1}{2}
  \int \left< P^0[h^F]_{\mu \nu}(x) \, h^F_{\rho \sigma}(y) \right>_0
  \, {P^\infty}^\dagger[t]^{\rho \sigma}(y) \, d^Ny +
  \nonumber \\
  & & \qquad \qquad \qquad + \frac{1}{2} \int \left<
  P^\infty[h^F]_{\mu \nu}(x)
  \, h^F_{\rho \sigma}(y) \right>_0
  \, {P^0}^\dagger[t]^{\rho \sigma}(y) \, d^Ny
\label{jds}
\end{eqnarray}
and using the fact that ${P^\infty}^\dagger$ and ${P^0}^\dagger$ act like
the identity on conserved sources we have that the l.h.s. of (\ref{jds})
is given by $P^S[h^F]_{\mu \nu}(x)$ being $h^F_{\mu \nu}$ the field,
computed in the harmonic gauge, associated to the conserved source
$t^{\rho \sigma}$. For $r \rightarrow \infty$, $h^0_{\mu \nu}(x)$
behaves from (\ref{Delta3}) like
\begin{eqnarray}
  h^0_{\mu \nu}(x) & = & - \frac{2 x^\alpha x^\beta}{r^2} \int_0^r dr' \,
  r' \, R^L_{\alpha \mu , \beta \nu}(r' \hat{x}) +
  \frac{2 x^\alpha x^\beta}{r^3} \int_0^r dr' \, {r'}^2
  R^L_{\alpha \mu , \beta \nu}(r' \hat{x}) = \nonumber \\
  & = & K^0_{\mu \nu}(x) + \frac{H^0_{\mu \nu}(x)}{r}
  + O(r^{-1-\epsilon})
\end{eqnarray}
while  $h^\infty_{\mu \nu}(x)$ from (\ref{Delta4}) is given by
\begin{equation}
  h^\infty_{\mu \nu}(x) = - \frac{2 x^\alpha x^\beta}{r^2}
  \int_r^\infty dr' \, r' \, R^L_{\alpha \mu , \beta \nu}(r' \hat{x}) +
  \frac{2 x^\alpha x^\beta}{r^3} \int_r^\infty dr' \, {r'}^2
  R^L_{\alpha \mu , \beta \nu}(r' \hat{x})
\end{equation}
and is regular in the same limit, in the sense that
it vanishes at least as quickly as $h^F_{\mu \nu}$
itself. Similarly in the limit $r \rightarrow 0$
we find that  $h^0_{\mu \nu}(x)$
vanishes and
\begin{equation}
  h^\infty_{\mu \nu}(x) = - K^0_{\mu \nu}(x) - \frac{H^0_{\mu \nu}(x)}{r}
  + O(r^\varepsilon),
\end{equation}
from which one obtains the two relations (\ref{lsu}) and (\ref{kdj}).
In Appendix A it is shown how one can express the propagators
$G^S$ in terms of the familiar hypergeometric functions, while in Appendix
B  we treat the $\beta\neq 0$ (i.e. non sharp) radial gauge in the second
order formalism.

\section{First order formalism}

\subsection{Radial gauge}

We recall for completeness the main formulae for the radial gauge in the
first order formalism given by \cite{M.T.}. The gauge is defined by
\begin{equation}
\xi^\mu\Gamma^{ab}_\mu(\xi)=0
\end{equation}
\begin{equation}
\xi^\mu e^{a}_\mu(\xi)=\xi^\mu \delta^a_\mu
\end{equation}
and $\Gamma^{ab}_\mu (\xi)$ and $e^a_{\mu}(\xi)$ are expressed in term of
the radial Riemann and torsion two forms as follows
\begin{equation}
\label{eqrad1}
\Gamma^{a}_{b,\mu}(\xi)=\xi^\nu\int^1_0 d\lambda ~\lambda
R^{a}_{b,\nu\mu}(\lambda\xi)
\end{equation}
\begin{equation}
\label{eqrad2}
e^{a}_{\mu}(\xi)=\delta^a_\mu+\xi^\nu\xi^b\int^1_0 d\lambda ~\lambda
(1-\lambda) R^{a}_{b,\nu\mu}(\lambda\xi)
+\xi^\nu \int^1_0 d\lambda ~\lambda S^{a}_{\nu\mu}(\lambda\xi)
\end{equation}
for the derivation of which we refer to \cite{M.T.}. These formulae hold
under the
hypothesis of regularity at the origin for the field $(\Gamma^{ab}_\mu (x),
e^a_\mu (x))$. Radial projections corresponding to different assumption about
the behavior of the field at the origin  will be considered in the following
subsection.
\subsection{ Radial Projectors}
Similarly as done in section 3, at the linearized level we can use relations
(\ref{eqrad1}) and (\ref{eqrad2})  to project the general field $(\G,e)$ into
the radial field $(\Gamma^R, e^R)$ as the r.h.s of eqs.(\ref{eqrad1}) and
(\ref{eqrad2}) are invariant under linearized gauge transformations
\begin{equation}
\left (
\begin{array}{c}
\Gamma^{ab}_{\mu}(x)\\
\\
\tau^{a}_{\mu}(x)
\end{array}
\right )
\longrightarrow
\left (
\begin{array}{c}
 \Gamma^{ab}_{\mu}(x)+\dev_\mu \Theta^{ab}(x)\\
\\
  \tau^{a}_{\mu}(x)-\Theta^{ab}(x)\e_{b\mu}+\dev_{\mu} \Lambda^a (x)
\end{array}
\right )
\end{equation}
where $\tau^{a}_\mu (x)$ is related to $e^a_\mu(x)$ by
\begin{equation}
e^a_\mu (x)=\delta^a_\mu+\tau^a_\mu (x).
\end{equation}
$\Theta^{ab}(x)$ is the infinitesimal Lorentz-transformation  and
$\Lambda^a(x)$ the infinitesimal translation. We shall be interested in the
projectors $P^{0}$ and $P^{\infty}$ to construct the projector
\begin{equation}
\label{1PS}
P^S=\frac{1}{2}(P^{0}+P^{\infty}).
\end{equation}
$P^0$ is given by (\ref{eqrad1}) and (\ref{eqrad2}) substituting to
$R^{ab}_{\mu\nu}$ and $S^a_{\mu\nu}$ their linearized expression
\begin{equation}
R^{L a b}_{\mu\nu}(x)=\partial_\mu \Gamma^{a b}_\nu (x) -\partial_\nu
\Gamma^{a b}_\mu (x)
\end{equation}
\begin{equation}
S^{L a }_{\mu\nu}(x)=\partial_\mu \tau^{a }_\nu (x) -\partial_\nu
\tau^{a}_\mu (x)
+\Gamma^{a b}_\mu (x)\e_{b\nu}-\Gamma^{a b}_\nu (x)\e_{b\mu}
\end{equation}
which are invariant under linearized gauge transformation; i.e.
\begin{equation}
\label{1P0}
P^{0}
\left (
\begin{array}{c}
\Gamma^{a}_{b,\mu}(x)\\
\\
\\
\tau^{a}_{\mu}(x)
\end{array}
\right )
=
\left (
\begin{array}{c}
x^\nu\displaystyle{\int}^1_0 d\lambda ~\lambda
R^{La}_{b,\nu\mu}(\lambda x)\\
\\
x^\nu x^b\displaystyle{\int}^1_0 d\lambda ~\lambda (1-\lambda)
R^{La}_{b,\nu\mu}(\lambda x)
+x^\nu \displaystyle{\int}^1_0 d\lambda ~\lambda S^{La}_{\nu\mu}(\lambda x)
\end{array}
\right )
\end{equation}

\begin{equation}
\label{1Pinfinity}
P^{\infty}
\left (
\begin{array}{c}
\Gamma^{a}_{b,\mu}(x)\\
\\
\\
\tau^{a}_{\mu}(x)
\end{array}
\right )
=
\left (
\begin{array}{c}
-x^\nu\displaystyle{\int}^\infty_1 d\lambda ~\lambda
R^{La}_{b,\nu\mu}(\lambda x)\\
\\
-x^\nu x^b\displaystyle{\int}^\infty_1 d\lambda ~\lambda (1-\lambda)
R^{La}_{b,\nu\mu}(\lambda x)
-x^\nu \displaystyle{\int}^\infty_1 d\lambda ~\lambda
S^{La}_{\nu\mu}(\lambda x)
\end{array}
\right ).
\end{equation}
 $P^0$ projects on radial fields that are regular at the origin (i.e.
behaving like the original field) and give a connection
$\Gamma^{ab}_\mu (x)$ behaving like ${1/r}$ at $\infty$.
On the other hand $P^\infty$ projects on a radial field that is regular
at $\infty$ and give a connection
$\Gamma^{ab}_\mu (x)$ behaving like ${1/r}$ at the
origin. As we noticed in the introduction the radial connection has to
behave like ${1/r}$ at the origin and/or at $\infty$,
otherwise the Wilson loop given by two radii and closed by two arcs one
going to the origin and the other going to infinity (see Fig.1) would be
identically 1, due to the radial nature of the connection;
but this would be contradictory as one cannot fix at the kinematical level the
value of a gauge invariant quantity. $P^S$ which allows us to
construct a finite propagator in $N>2$, treats the origin and infinity in
symmetrical way by giving the same ${1/r}$ behavior in
the two limits with opposite coefficients. This is seen in the same way as
shown at the end of section 3 and in Appendix A in the second order formalism.
Eq. (\ref{1PS}) together with eqs. (\ref{1P0}) and (\ref{1Pinfinity}) will
allow us to compute the radial propagator associated to the $P^S$
projector, in terms of the correlation functions of the linearized Riemann
and torsion two forms, which are invariant under linearized gauge
transformations and as such can be computed in any gauge (e.g. the harmonic
gauge). First we must derive the action of the adjoint projectors on the
torsion source $\Sigma^\mu_{ab}=-\Sigma^\mu_{ba}$ and on the energy
momentum tensor $T^\mu_a$.
\begin{eqnarray}
\label{P0dagS}
&&P^{0\dagger}(\Sigma)^\mu_{ab}=\Sigma^\mu_{ab}(x)+x^\mu \int^\infty_1
d\lambda \lambda^{N-1}\left (\partial_\rho \Sigma^\rho_{ab}(\lambda
x)+\frac{1}{2}\left (
T^\rho_a(\lambda x)\eta_{\rho b} -T^\rho_b(\lambda x)\eta_{\rho
a}\right)\right)\nonumber\\
&&+\frac{x^\mu}{2}\int^\infty_1 d\lambda
(\lambda^N-\lambda^{N-1})(x_b\partial_\rho T^\rho_a-x_a\partial_\rho T^\rho_b)
\end{eqnarray}
\begin{equation}
\label{P0dagT}
P^{0\dagger}(T)^\mu_a=T^\mu_a (x)+x^\mu\int^\infty_1 d\lambda \lambda^{N-1}
\partial_\rho T^\rho_a (\lambda x)
\end{equation}
The $P^{\infty\dagger}$ are given by changing in (\ref{P0dagS}) and
(\ref{P0dagT})  the integration limit from $(1,\infty)$ to $(1,0)$.
One easily verifies from (\ref{P0dagS}) and
(\ref{P0dagT})  that the projected sources $P^{0\dagger}(\Sigma,T)$ satisfy
the linearized Bianchi identities, as proved on general grounds in section 2
eq. (2.17)
\begin{equation}
\label{Bianchi1}
\partial_\mu \Sigma^\m_{ab}+\frac{1}{2}(T_{a}^\rho\eta_{\rho b}-T_{b}^\rho
\eta_{\rho a} )=0
\end{equation}
\begin{equation}
\label{Bianchi2}
\partial_\mu T^\mu_a=0.
\end{equation}
In addition one notices that the r.h.s of (\ref{P0dagS}) and
(\ref{P0dagT}) are integrals of the linearized Bianchi identities
(\ref{Bianchi1}) and (\ref{Bianchi2}) and thus $P^{0\dagger}$ and
$P^{\infty\dagger}$ leave unchanged the sources satisfying the Bianchi
identities, in agreement with the general argument of section 2.
\subsection{Radial Propagators}
Similarly to what happens in the second order formalism, the propagators
constructed from $\langle P^0(\Gamma,\tau) P^0(\Gamma,\tau)\rangle_0$
are divergent while $\langle P^\infty(\Gamma,\tau)
P^\infty(\Gamma,\tau)\rangle_0$  diverge for $N\le 4$. Thus we shall
construct
the solution for the the $P^S$ projected Green' s function equation
\begin{equation}
\label{EQS}
\left (
\begin{array}{ccc}
\frac{1}{2}(\delta_{am}^{\mu\nu}\eta_{bn}-\delta_{an}^{\mu\nu}\eta_{bm}) &
& \delta^{\lambda\mu\sigma}_{abd}\partial_\lambda\\
 & & \\
\delta^{\mu\lambda\nu}_{pmn}\partial_\lambda & & 0
\end{array}
\right)
\left (
\begin{array}{ccc}
G^{mn,rs}_{\nu,\gamma} & & G^{mn,g}_{\nu,\beta}\\
 & \\
G^{d,rs}_{\sigma,\gamma} & & G^{d,g}_{\sigma,\beta}
\end{array}
\right )
=
P^{S\dagger}\left (
\begin{array}{ccc}
\delta^\mu_\gamma \delta^{rs}_{ab} & &  0\\
 &  &\\
0 & &\delta^\mu_\beta \delta^g_p
\end{array}
\right )\delta^N (x-y)
\end{equation}
by means of general technique described in section 2 and 3. One proves that
\begin{eqnarray}
\label{1GPS}
&&-i G^S (x,y)=\\
&&\frac{1}{2}\left \langle P^{0}
\left (
\begin{array}{c}
\G^{a}_{b,\m}(x)\\
\\
\t^{a}_{\m}(x)
\end{array}
\right )
P^{\infty}
\left (
\G^{a}_{b,\m}(y)\;
\t^{a}_{\m}(y)
\right )\right \rangle
+\frac{1}{2}
\left \langle P^{\infty}
\left (
\begin{array}{c}
\G^{a}_{b,\m}(x)\\
\\
\t^{a}_{\m}(x)
\end{array}
\right )
P^{0}
\left (
\G^{a}_{b,\m}(y)\;
\t^{a}_{\m}(y)
\right )\right \rangle\nonumber
\end{eqnarray}
is a convergent symmetric radial solution of (\ref{EQS}) for all
$N>2$. The explicit form of solution (\ref{1GPS}) of the Green' s function
equation (\ref{EQS})  is easily computed  by using (\ref{1P0}) and
(\ref{1Pinfinity})  where the correlators between Riemann and torsion two-forms
which are invariant under  linearized gauge transformations, can be
obtained using e.g. the usual symmetric harmonic gauge ($\beta=\infty$) in
which
\begin{equation}
\label{TT}
\langle
\tau_{a\mu}(x)\tau_{b\nu}(y)\rangle=-i\left(\eta_{ab}\eta_{\mu\nu}+\eta_{a\nu}
\eta_{b\mu}-\frac{2}{N-2}\eta_{a\mu}\eta_{b\nu}\right)D(x-y).
\end{equation}
We get
\begin{equation}
\label{RR}
\langle R^{Lab}_{\mu\nu}(x) R^{Lcd}_{\lambda\rho}(y)\rangle=
\langle R^{Lab}_{\mu\nu}(x) R^{Lcd}_{\lambda\rho}(y)\rangle^{II}+
[(\partial_{\mu}\partial_{\lambda} M^{ab,cd}_{\nu,\rho}-
\partial_{\mu}\partial_{\rho} M^{ab,cd}_{\nu,\lambda})-
(\mu\leftrightarrow\nu)]
\end{equation}
where $\langle\ \rangle^{II} $ means the correlator in the second order
formalism (see section 4) and
\begin{equation}
\label{GG}
 M^{ab,cd}_{\nu,\rho}=-\frac{i}{4}\left(
\delta_{\nu\rho}\delta^{ab}_{c'd'}\eta^{c'c}\eta^{d'd}
+\left (\delta^{ab}_{\nu\alpha}
\delta^{cd}_{\rho\beta}-\frac{2}{N-2}\delta^{ab}_{\rho\alpha}
\delta^{cd}_{\nu\beta}\right )\eta^{\alpha\beta}\right)\delta^N(x-y)
\end{equation}
is the usual ultralocal part of the propagator of $\Gamma$ in the first
order formalism \cite{M.P.}.
\begin{equation}
\label{SS}
\langle S^{La}_{\mu\nu}(x) S^{Lb}{\rho\sigma}(y)\rangle=
[ (M^{a c,b d}_{\mu,\rho}\eta_{c\nu}\eta_{d\sigma}-
M^{a c,b d}_{\mu,\sigma}\eta_{c\nu}\eta_{d\rho})-
(\mu\leftrightarrow\nu)]
\end{equation}
\begin{equation}
\label{RS}
\langle R^{Lab}_{\mu\nu}(x) S^{Lc}_{\rho\sigma}(y)\rangle=
[(\partial_{\mu} M^{ab,cd}_{\nu,\rho}\eta_{d\sigma}-
\partial_{\nu} M^{ab,cd}_{\mu,\rho}\eta_{d\sigma})-
(\rho\leftrightarrow\sigma)
].
\end{equation}
The local nature of the correlators (\ref{SS}) and (\ref{RS}) reflects the
well known fact that the torsion does not propagate in the Einstein-Cartan
theory.
The propagators are obtained by substituting the correlators of the Riemann
and torsion tensors given by (\ref{RR}), (\ref{SS}) and (\ref{RS}) into
(\ref{1GPS}).
The integrals over
 $\lambda$ and $\tau$ of the $\langle R R \rangle^{II}$ correlators are the
same as those given in the second order formalism.
The contact terms generated by the expression of $M^{ab,cd}_{\nu,\rho}$
and appearing on the r.h.s. eqs. (\ref{RR}),(\ref{SS}) and (\ref{RS})
lead to integrals in $\lambda$ and $\tau$ which are
convergent. In fact the generic integral is of the form
\begin{eqnarray}
&&\int^1_0 d\lambda \lambda^A \int^\infty_1 d\tau \tau^B
\delta^N(\lambda x-\tau y)=\nonumber \\
&&\int^1_0 d\lambda \lambda^A \int^\infty_1 d\tau \tau^B \delta (\lambda
|x|-
\tau |y|)(\lambda |x|)^{-N+1} \delta^{N-1}(\Omega_x-\Omega_y)=\\
&&\delta^{N-1}(\Omega_x-\Omega_y) \Theta (|x|-|y|)\frac{1}{A+B-N+2}
\left ( |y|^{-B-1}|x|^{B-N+1}-|y|^{A-N+1}|x|^{-A-1} \right ).\nonumber
\end{eqnarray}
We notice that in three dimensions the Riemann-Riemann correlator (\ref{RR}) is
identically zero
as can be
explicitly verified substituting (\ref{TT}) and (\ref{GG}) into (\ref{RR}) or
alternatively,
observing
that the (2,1)
component of eq. (\ref{EQS}) in three dimensions reads  $\langle
R^L(x)\Gamma(y)\rangle
\equiv 0 $ which implies that $\langle R^L(x) R^L(y)\rangle\equiv 0 $ .
This means
through (\ref{1P0}) and (\ref{1Pinfinity}) that
$\langle\Gamma(x)\Gamma(y)\rangle\equiv 0 $
and that
$\langle
\tau(x)
\tau(y)\rangle $ is zero except for singular contributions that arise when
the
origin,
$x$ and $y$ are collinear; such contributions are given by (\ref{RR}),
(\ref{SS}) and (\ref{RS}).
Also in the second order approach the correlator $\langle h(x)h(y)\rangle$
 in three
dimensions reduces to the collinear singular terms. Thus in the radial gauge
one
reads
directly the non existence of propagating gravitons, while in the harmonic
gauge
one
propagates a pure gauge.
The surviving collinear singularity of the radial propagator in three
dimension
reflects
the topological nature of the theory.

\section{Conclusions}
In the present paper we considered the quantization of the gravitational
field  in the radial gauge.
The main advantage of this gauge is to provide physical correlation
functions,
i.e.
correlation functions defined at points described invariantly in terms of a
geodesic
system of coordinates radiating from a given origin.
Given a generic field we developed general projection formulae for
extracting
from it a radial field that is gauge equivalent to the given one; the
projection
is obviously not unique due  to the presence of the residual gauge. For
reasons
intrinsic
to the gauge nature of the considered theory and that can be easily
understood
by
considering a special family of Wilson loops, such projected fields develop
some
kind
of singular behavior at the origin or a very slow decrease at infinity. The
best
choice is to treat the origin and infinity in symmetrical way by sharing
equally
such non regular behavior and this fixes completely the gauge.  Technically
the
projection procedure
is obtained through
the Riemann and torsion tensors which are invariant at the linearized level.
Still
the direct computation of the v.e.v. of the symmetrically projected field
contains
an infinite solution of the homogeneous radially projected Green's
function, due to the singular behavior of the Riemann-Riemann  correlator,
which has to be subtracted away. In this way one obtains an explicit
solution of
the
radially projected Green's function equation that is symmetric in the field
arguments,
radial and finite for all dimensions larger than 2.
In three dimensions the propagators vanish identically except for collinear
contributions; this is in agreement with the absence of propagating
gravitons in three dimensions, while the collinear contributions are
remnants at the perturbative level of topological effects
(conical defects) introduced by matter
in three dimensional gravity \cite{Flatland}.

\appendix{}

In this Appendix we show how the radial propagators discussed in the text
can be written in terms of the hypergeometric functions. We
start for simplicity sake from the case of Yang-Mills theory or quantum
electrodynamics in the euclidean formulation,
in which the propagator is given by \cite{M.S.}
\begin{equation}
  G^S_{\mu \nu}(x,y) = \frac{1}{2} x^\rho y^\sigma
  D_{\mu \rho \nu \sigma} [D(x,y) + D(y,x)]
\end{equation}
with
\begin{equation}
  D_{\mu \rho \nu \sigma} = \delta^{\mu' \rho'}_{\mu \rho} \,
  \delta^{\nu' \sigma'}_{\nu \sigma} \, \delta_{\rho' \sigma'}
  \, \frac{\partial}{\partial x^{\mu'}}
  \frac{\partial}{\partial y^{\nu'}}
\end{equation}
and
\begin{equation}
  D(x,y) = \frac{\Gamma(\frac{N}{2}-1)}{4 \pi^{N/2}} \,
  \int_0^1 d\lambda \int_1^\infty d\tau \,
  \frac{1}{[(\lambda x - \tau y)^2]^{\frac{N}{2}-1}} .
  \label{kky}
\end{equation}
$D(x,y)$ can be easily rewritten performing the change of variable
$\rho=\lambda/\tau$ and then integrating in $\lambda$, as follows
\begin{equation}
  D(x,y) = - \frac{\Gamma(\frac{N}{2}-1)}{4 \pi^{N/2}} \,
  \frac{1}{N-4} \int_0^1 d\rho \,
  \frac{1}{[(\rho x - y)^2]^{\frac{N}{2}-1}} \, [\rho^{N-4} - 1] .
\end{equation}
Writing
\begin{equation}
  \int_0^1 \, d\rho \frac{1}{[(\rho x - y)^2]^{\frac{N}{2}-1}} =
  \int_0^\infty d\rho \, \frac{1}{[(\rho x - y)^2]^{\frac{N}{2}-1}} -
  \int_0^1 d\rho \, \frac{\rho^{N-4}}{[(x - \rho y)^2]^{\frac{N}{2}-1}}
\end{equation}
we obtain
\begin{eqnarray}
  D(x,y) & = & - \frac{\Gamma(\frac{N}{2}-1)}{4 \pi^{N/2}} \,
  \frac{1}{N-4} \left\{ \int_0^1 d\rho \, \rho^{N-4}
  \left[ \frac{1}{[(\rho x - y)^2]^{\frac{N}{2}-1}} +
  \right. \right. \nonumber \\
  & & \left. \left. + \frac{1}{[(x - \rho y)^2]^{\frac{N}{2}-1}}
  \right]
  - \int_0^\infty d\rho \, \frac{1}{[(\rho x - vy)^2]^{\frac{N}{2}-1}}
  \right\}
\label{dxe}
\end{eqnarray}
\noindent The first integral appearing in (\ref{dxe}) can be rewritten as
follows
\begin{equation}
\label{I1}
I_1=\int^1_0~d\rho\frac{\rho^{N-4}}{[\rho^2 X^2-2 \rho X Y \cos\theta
+Y^2]^{N/2-1}}
\end{equation}
where $X=|x|$, $Y=|y|$ and performing the change of variable
$\displaystyle{\rho=\frac{\frac{Y}{d}\zeta}{\frac{Y}{d}\zeta+1}}$ with
$d=\sqrt{X^2+Y^2-2 X Y \cos\theta}$~ one obtains \cite{BE}
\begin{eqnarray}
&&Y^{-1}d^{3-N}\int^\infty_0 d\zeta\frac{\zeta^{N-4}}{\left [\zeta^2+2 \zeta
\left(\frac{ Y-X\cos\theta}{d}\right )+1\right ]^{N/2-1}}=\nonumber\\
&&\nonumber\\
&&=\frac{1}{N-3}Y^{-1}d^{3-N}
\left (4\sin^2{\phi_x}\right)^{\frac{3-N}{4}}  \Gamma\left (
\frac{N-1}{2}\right
){\rm P}^{\frac{3-N}{2}}_{\frac{3-N}{2}}\left (\cos{\phi_x}\right)
\end{eqnarray}
being $\displaystyle{\cos\phi_x=\frac{ Y-X\cos\theta}{d}}$. While
$ \theta$ is the angle between  $x$ and $y$, $\phi_x$ is the angle
opposite to $x$ in the euclidean triangle of sides $x,~y,~d$. Expressing
${\rm P}^{\frac{3-N}{2}}_{\frac{3-N}{2}}$ in terms of hypergeometric functions
\cite{BE} , we have
\begin{equation}
{\label{I11}}
I_1=
 \frac{1}{N-3} \, {d^{3-N}}{Y^{-1}}
\left(\cos^2\frac{\phi_x}{2}\right )^{\frac{3-N}{2}}
 {}_2 F_1 \left( \frac{N-3}{2} , \frac{5-N}{2} ; \frac{N-1}{2} ;
  \sin^2\frac{\phi_x}{2} \right).
\end{equation}
Similarly one computes
\begin{equation}
\label{I2}
I_2=
\int^\infty_0~d\rho\frac{1}{[\rho^2 X^2-2 \rho X Y \cos\theta
+Y^2]^{N/2-1}}
\end{equation}
to obtain
\begin{equation}
\label{I22}
I_2=\frac{1}{N-3} \, {Y^{3-N}}{X^{-1}}
\left(\sin^2\frac{\theta}{2}\right )^{\frac{3-N}{2}}
  {\ }_2 F_1 \left( \frac{N-3}{2} , \frac{5-N}{2} ; \frac{N-1}{2} ;
 \cos^2\frac{\theta}{2} \right).
\end{equation}
The behavior of $D(x,y)+D(y,x)$ for $X\to 0$, $y$ and $\theta$ fixed, is
obtained from $I_1$ and $I_2$ keeping in mind that in such limit $\phi_x\to
0$ and $\phi_y\to \pi-\theta$ thus giving
\begin{eqnarray}
&D(x,y)+D(y,x)\simeq\nonumber \\
&\frac{1}{N-3} \,
\left(\sin^2\frac{\theta}{2}\right )^{\frac{3-N}{2}}
 {\ }_2 F_1 \left( \frac{N-3}{2} , \frac{5-N}{2} ; \frac{N-1}{2} ;
 \cos^2\frac{\theta}{2}
\right)({Y^{3-N}}{X^{-1}}-{X^{3-N}}{Y^{-1}}).
\end{eqnarray}
This is also the behavior for $x$ and $\theta$ fixed and $Y\to\infty$. Due
to the symmetry of $D(x,y)+D(y,x)$ we have that for $Y\to 0$, $x$ and
$\theta$ fixed
\begin{eqnarray}
&D(x,y)+D(y,x)\simeq\nonumber\\
&-\frac{1}{N-3} \,
\left(\sin^2\frac{\theta}{2}\right )^{\frac{3-N}{2}}
 {\ }_2 F_1 \left( \frac{N-3}{2} , \frac{5-N}{2} ; \frac{N-1}{2} ;
 \cos^2\frac{\theta}{2}
\right)({Y^{3-N}}{X^{-1}}-{X^{3-N}}{Y^{-1}})
\end{eqnarray}
and also for $X\to \infty$, $y$ and $\theta$ fixed. We see that the
behavior for  $X\to \infty$ is
just the opposite of that for $X\to 0$. Being $D_{\mu\nu\rho\sigma}$ a zero
degree operator the same holds for the propagator $G^S_{\mu\nu}(x,y)$.
The addition to the computed propagator of a residual gauge
term, i.e. $\displaystyle{\frac{\partial F_\nu(x,y)}{\partial x^\mu}}$
with
\begin{equation}
\label{radiality}
\displaystyle{x^\mu\frac{\partial F_\nu(x,y)}{\partial x^\mu}}=0
\end{equation}
which obviously satisfies the homogenous equation
\begin{equation}
\displaystyle{\left (
\Box_x \delta_{\mu\alpha}-{\frac{\partial }{\partial x^\mu}}{\frac{\partial
}{\partial x^\alpha}}\right ) {\frac{\partial F_\nu(x,y)}{\partial x^\alpha}}
=0}
\end{equation}
 destroys such a opposite behavior. In fact eq. (\ref{radiality}) tells us
that $F_\nu (x,y)$ is a homogeneous function of $x$ of degree zero and as
such $\displaystyle{\frac{\partial F_\nu(x,y)}{\partial x^\mu}}$ behaves
for $X\to 0 $ and $X\to\infty$ like $a(\theta ,y)/X$ with the same
coefficient $a(\theta,y)$. Thus the imposition of the opposite  behavior at
the origin and infinity fixes the radial gauge completely.

In the case of gravity the integrals of the text can be written
as combinations of
\begin{equation}
  \int_0^1 d\rho \, \frac{\rho^{c-b-2}}{[(\rho x - y)^2]^\frac{c}{2}}
\label{hsv}
\end{equation}
and
\begin{equation}
  \int_0^\infty d\rho \, \frac{\rho^a}{[(\rho x - y)^2]^\frac{c}{2}}
\label{jdc}
\end{equation}
with $c=N$ and where $a$ and $b$ take the values 0 or 1; or with
$c=N+2$ where $a$ and $b$ take the values 0,1,2,3. Eq.(\ref{jdc}) is
 treated as (\ref{I2}) \cite{BE}, while (\ref{hsv}) can be written
as a combination of hypergeometric functions  using \cite{BE}
\begin{eqnarray}
  &&\int_0^1 d\rho \,
  \frac{\rho^{c-n-2}(1-\rho)^n}{[(\rho x - y)^2]^\frac{c}{2}} =
  \nonumber \\
  && ={d^{1+n-c}}{Y^{-1-n}}\int^\infty_0 d\zeta
   \frac{\zeta^{c-n-2}}{[\zeta^2 +2 \zeta \cos\phi_x
+1]^\frac{c}{2}}={\rm B}(n+1,c-n-1)\times\nonumber\\
   && {d^{1+n-c}}{Y^{-1-n}}
 \left (\cos^2\frac{\phi_x}{2}\right )^{\frac{1-c}{2}}
{\ }_2 F_1 \left( \frac{c-2 n-1}{2} , \frac{2 n+3-c}{2} ; \frac{c+1}{2} ;
  \sin^2 \frac{\phi_x}{2} \right) .
\end{eqnarray}
With a method similar to that explained above one proves that the behavior
of
\begin{equation}
{\label F}
F^S_{\alpha\beta}(x,y)+F^S_{\beta\alpha}(y,x)
\end{equation}
for $X\to 0$ is of the form
\begin{equation}
f_0(x,y)(X^{-2}Y^{-N+2}-Y^{-2}X^{-N+2})+
g_0(x,y)(X^{-1}Y^{-N+1}-Y^{-1}X^{-N+1})
\end{equation}
and for $X\to \infty$
\begin{equation}
-f_0(x,y)(X^{-2}Y^{-N+2}-Y^{-2}X^{-N+2})
-g_0(x,y)(X^{-1}Y^{-N+1}-Y^{-1}X^{-N+1})
\end{equation}
where $f_0$ and $g_0$ are homogeneous functions of degree zero in $x$ and
$y$.  One
reaches the propagator by applying to ({\ref F}) the operator obtained from
(\ref{RR2}) and (\ref{FS2})
\begin{eqnarray}
  & & D_{\nu ,\beta;\sigma,\gamma;\alpha^\prime \lambda^\prime}=
  \frac{1}{4} x^\mu x^\alpha y^\rho y^\lambda
   \delta^{\mu' \nu'}_{\mu \nu} \delta_{\nu' \nu''} \, (
  \delta^{\rho' \nu''}_{\rho \sigma} \delta^{\alpha' \beta'}_{\alpha \beta}
  \delta^{\lambda' \beta''}_{\lambda \gamma}
  \delta_{\beta' \beta''}+ \nonumber \\
  & & \qquad + \, \delta^{\lambda' \nu''}_{\lambda \gamma}
  \delta^{\alpha' \beta'}_{\alpha \beta} \delta^{\rho' \beta''}_{\rho \sigma}
  \delta_{\beta' \beta''} -
  \mbox{$\frac{2}{N-2}$} \,
  \delta^{\alpha' \nu''}_{\alpha \beta} \delta^{\rho' \sigma'}_{\rho \sigma}
  \delta^{\lambda' \sigma''}_{\lambda \gamma}
  \delta_{\sigma' \sigma''} ) \
  \partial_{\mu'}  \partial_{\rho'}
\end{eqnarray}
Being this an operator of dimension 1 in $x$ and $y$ it
generates a propagator behaving  for $X\to 0$ like
\begin{equation}
\tilde f_0(x,y)(X^{-1}Y^{-N+3}-Y^{-1}X^{-N+3})+
\tilde g_0(x,y)(Y^{-N+2}-X^{-N+2})
\end{equation}
and the opposite for $X\to\infty$. Similarly to what happens in the Yang-Mills
case, addition of a residual gauge term destroys the opposite behavior
between the origin and infinity. In fact a residual gauge term has the
form
\begin{equation}
\label{ff}
\frac{\partial F_{\nu,\alpha\beta}(x,y)}{\partial x^\mu}+
\frac{\partial F_{\mu,\alpha\beta}(x,y)}{\partial x^\nu}
\end{equation}
with
\begin{equation}
x^\mu\frac{\partial F_{\nu,\alpha\beta}(x,y)}{\partial x^\mu}+
x^\mu\frac{\partial F_{\mu,\alpha\beta}(x,y)}{\partial x^\nu}=0.
\end{equation}
Multiplying by $x^\nu$ one easily proves that $x^\nu
F_{\nu,\alpha\beta}(x,y)$ is a homogeneous function of degree 1 in $x$
 and using this fact one shows that
\begin{equation}
 F_{\nu,\alpha\beta}(x,y)=H^0_{\nu,\alpha\beta}(x,y)+H^1_{\nu,\alpha\beta}(x,y)
\end{equation}
being $H^0$ and $H^1$ homogenous functions of $x$ of degree 0 and 1.
Thus (\ref{ff}) has the same $a(\theta,y)/X+b(\theta,y)$ behavior at the
origin and infinity thus violating the opposite behavior of the propagator.

\appendix{}

In the main text we computed the propagator in the ``sharp'' radial gauge
both in the first and in the second order formalism. Now we give a general
technique to derive the propagator in presence of a radial gauge-fixing term
in the lagrangian. As an example we consider the second order formalism with
the following gauge-fixed lagrangian
\begin{equation}
\label{lagrangian}
{\cal L}(x)={\cal L}^{II}(x)+\frac{1}{2\beta}
(x^\mu h_{\mu\nu}(x) x^\lambda h_{\lambda}^{\nu}(x)).
\end{equation}
The sharp case can be recovered in the limit $\beta\to 0$. The addition of
the  term in $\beta$ modifies the equation for  the propagator
in the following way
\begin{equation}
K_{x\mu\nu}^{\rho\sigma} G_{\rho\sigma,\alpha\beta}(x,y)+K_{x\mu\nu}^{R
\rho\sigma }
G_{\rho\sigma,\alpha\beta}(x,y)=
\frac{1}{2}(\eta_{\mu\alpha}\eta_{\nu\beta}+\eta_{\mu\beta}\eta_{\nu\alpha})
\delta^N (x-y)
\end{equation}
where $K^{\rho\sigma}_{x\mu\nu}$ is the usual kinematical term defined in
the subsection 2.C and $K_{x\mu\nu}^{R\rho\sigma}$ is given by
\begin{equation}
K_{x\mu\nu}^{R\rho\sigma}=\frac{1}{4\beta}(x_\mu x^\rho \delta^\sigma_\nu +
x_\nu x^\sigma \delta^\rho_\mu+ x_\mu x^\sigma \delta^\rho_\nu+x_\nu x^\rho
\delta^\sigma_\mu).
\end{equation}
We shall solve this equation writing $G=G^{(1)}+G^{(2)}$ with
\begin{equation}
\label{X1}
K_{x\mu\nu}^{R\rho\sigma}G^{(1)}_{\alpha\beta ,\rho\sigma}(x,y)=
-P^\dagger\left
 (\frac{1}{2}(\eta_{\mu\alpha}
\eta_{\nu\beta}+\eta_{\mu\beta}
\eta_{\nu\alpha})\delta^N (x-y)\right ),
\end{equation}
\begin{equation}
\label{X2}
K_{x\mu\nu}^{R\rho\sigma} G^{(2)}_{\alpha\beta ,\rho\sigma}(x,y)=-(1-P^\dagger)
\left (\frac{1}{2}
(\eta_{\mu\alpha}\eta_{\nu\beta}+\eta_{\mu\beta}\eta_{\nu\alpha})\delta^N (x-y)
\right )
\end{equation}
and at the same time
\begin{equation}
\label{Y1}
K_{x\mu\nu}^{R\rho\sigma}G^{(1)}_{\alpha\beta ,\rho\sigma}(x,y)=0,
\end{equation}
\begin{equation}
\label{Y2}
K_{x\mu\nu}^{\rho\sigma}G^{(2)}_{\alpha\beta,\rho\sigma}(x,y)=0.
\end{equation}
We know already a solution of eq.(\ref{X1}) and eq.(\ref{Y1}). In fact the
``sharp'' radial propagators that we  found in the paper satisfy both
equations. We look now for a solution of the system eq.(\ref{X2}) and
eq.(\ref{Y2}). If $G^{(2)}$ is a pure gauge term, i.e.
\begin{equation}
\label{Gauge}
\frac{\partial}{\partial x^\mu} F_{\nu\alpha\beta}(x,y)+
\frac{\partial}{\partial x^\nu} F_{\mu\alpha\beta}(x,y),
\end{equation}
it automatically solves eq.(\ref{Y2}).
Now substituting (\ref{Gauge}) into eq.(\ref{X2}) we find the following
equation
\begin{eqnarray}
\label{AA}
&&\frac{1}{4\beta}\left (x_\mu x^\rho \left (\frac{\partial}{\partial x^\rho}
F_{\nu\alpha\beta}(x,y)+
\frac{\partial}{\partial x^\nu} F_{\rho\alpha\beta}(x,y)\right)+
x_\nu x^\rho \left (\frac{\partial}{\partial x^\rho}
F_{\mu\alpha\beta}(x,y)+
\frac{\partial}{\partial x^\mu}
F_{\rho\alpha\beta}(x,y)\right)\right)=\nonumber\\
&&=-(1-P^\dagger)\left (\frac{1}{2}
(\eta_{\mu\alpha}\eta_{\nu\beta}+\eta_{\mu\beta}\eta_{\nu\alpha})\delta^N (x-y)
\right ).
\end{eqnarray}
Using the explicit form of the projectors one can see that the
tensor structure of the r.h.s. of eq.(\ref{AA}) is the following
\begin{equation}
x_{\mu}j_{\nu,\alpha\beta}(x,y) +x_{\nu}j_{\mu,\alpha\beta}(x,y)
\end{equation}
where $j_{\mu,\alpha\beta}(x,y)$ changes according to the projector chosen.
Thus equating the same tensor structures we obtain
\begin{equation}
\label{DD}
\frac{x^\rho}{4\beta} \left (\frac{\partial}{\partial x^\rho}
F_{\nu\alpha\beta}(x,y)+
\frac{\partial}{\partial x^\nu} F_{\rho\alpha\beta}(x,y)\right)=
j_{\nu,\alpha\beta}(x,y).
\end{equation}
This equation can be solved with the usual technique  described in the
paper. We
notice  that the inversion of the dilatator that appears in (\ref{DD}) has
to be
done taking in account the boundary condition that the propagator must
satisfy. In this way the solution is automatically symmetric. For example
if we choose the case of $P^{\infty}$ the solution is given by
\begin{equation}
\frac{1}{4\beta} F^\infty_{\nu,\alpha\beta}(x,y)=-\int_1^\infty
\frac{d\alpha}{\alpha^2}
j^{\infty}_{\nu,\alpha\beta}(\alpha x,y)
-\frac{1}{2}\frac{\partial}{\partial x^\nu}
\int_1^\infty \frac{d\alpha}{\alpha}\left (1-\frac{1}{\alpha}\right)
x^\rho j^{\infty}_{\rho,\alpha\beta}(\alpha x,y)
\end{equation}
 where
\begin{eqnarray}
&&j^{\infty}_{\nu,\alpha\beta}(x,y)=-
\frac{1}{2}\eta_{\nu\beta}\frac{\partial}{\partial y^\alpha}
\int_1^\infty \frac{d\alpha}{\alpha^2} \delta(x-\alpha y)
-\frac{1}{2}\eta_{\nu\alpha}\frac{\partial}{\partial y^\beta}
\int_1^\infty \frac{d\alpha}{\alpha^2} \delta(x-\alpha y)\nonumber\\
&&-\frac{1}{2}\frac{\partial}{\partial y^\alpha}
\frac{\partial}{\partial y^\beta}
\int_1^\infty \frac{d\alpha}{\alpha^2}(\alpha-1)y_\nu \delta(x-\alpha y).
\end{eqnarray}
The final result is
\begin{eqnarray}
\label{Ris}
&&\frac{1}{4\beta} F^{\infty}_{\nu,\alpha\beta}(x,y)=
\frac{1}{2}\eta_{\nu\beta}\frac{\partial}{\partial y^\alpha}
\int_1^\infty \frac{d\alpha}{\alpha^2}\int_1^\infty \frac{d\lambda}{\lambda^2}
 \delta(\lambda x-\alpha y)+
\frac{1}{2}\eta_{\nu\alpha}\frac{\partial}{\partial y^\beta}
\int_1^\infty \frac{d\alpha}{\alpha^2}\int_1^\infty \frac{d\lambda}{\lambda^2}
 \delta(\lambda x-\alpha y)
\nonumber\\
&&+\frac{1}{2}\frac{\partial}{\partial y^\alpha}\frac{\partial}{\partial
y^\beta}
\int_1^\infty \frac{d\alpha}{\alpha^2}\int_1^\infty \frac{d\lambda}{\lambda^2}
(\alpha-1) y_\nu \delta(\lambda x-\alpha y)+\nonumber\\
&&+\frac{1}{4}\frac{\partial}{\partial x^\nu}\biggl (x_\beta
\frac{\partial}{\partial y^\alpha}
\int_1^\infty \frac{d\alpha}{\alpha^2}\int_1^\infty \frac{d\lambda}{\lambda^2}
(\lambda-1)  \delta(\lambda x-\alpha y)\nonumber+\\
&&+x_\alpha
\frac{\partial}{\partial y^\beta}
\int_1^\infty \frac{d\alpha}{\alpha^2}\int_1^\infty \frac{d\lambda}{\lambda^2}
(\lambda-1)  \delta(\lambda x-\alpha y)\biggr )\nonumber\\
&&-\frac{1}{4}\frac{\partial}{\partial x^\nu}   \frac{\partial}{\partial
y^\alpha}       \frac{\partial}{\partial y^\beta}
\int_1^\infty \frac{d\alpha}{\alpha^2}\int_1^\infty \frac{d\lambda}{\lambda^2}
(\lambda-1)(\alpha-1) (x\cdot y) \delta(\lambda x-\alpha y)
\end{eqnarray}
We have not performed the integrals to  enlighten the
symmetry of the result between $x$ and $y$ when one substitutes eq.(\ref{Ris})
into  eq.(\ref{Gauge}). The complete integration of the previous expression
can be executed with the help of the general formula given at the end of
section 4.
The case $P^S$ can be treated similarly. The generalization to the first
order formalism is trivial. For Yang-Mills theory see \cite{M.S.}.

\newpage
\noindent{\Large\bf Figure Captions}\\

\noindent Fig.1: Wilson loop restricting the behavior of the connection at
the origin and infinity\\

\end{document}